\documentclass[aps,prb,floatfix,twocolumn,superscriptaddress,showpacs,epsf]{revtex4-1}

\usepackage[USenglish]{babel}
\usepackage{amsmath,amssymb,amsfonts}
\usepackage{epstopdf}
\usepackage{epsfig} 
\usepackage{graphicx}
\usepackage{subfigure}
\usepackage{import}
\usepackage{color}
\usepackage{url}
\usepackage{changes}

\usepackage{enumitem}

\allowdisplaybreaks

\usepackage[colorlinks=true,linkcolor=blue,citecolor=red]{hyperref}

\def\bcen{\begin{center}}
\def\ecen{\end{center}}

\def\a{\alpha}       \def\b{\beta}      \def\d{\delta}

                    \def\s{\sigma}

       \def\D{\Delta}    \def\L{\Lambda}

\def\=={\equiv}

\def\qed{\raise1pt\hbox{\vrule height5pt width5pt depth0pt}}

\def\cG0{{\cal G}_0} 
\def\cG{{\cal G}}

\def\up{\uparrow} \def\down{\downarrow} 

\def\bk{{\bf k}}

 \def\=={\equiv}

 \def\ep0{\epsilon_{p}} \def\ed0{\epsilon_{f}}

\def\be{\begin{equation}}
\def\ee{\end{equation}}

\def\cc{c^{\dagger}}
\def\ca{c^{\phantom{\dagger}}}

\newcommand{\bd}[1]{\mathbf{#1}}

\begin{document}

\title{Interface and bulk superconductivity in superconducting heterostructures with enhanced critical temperatures}
\author{Giacomo Mazza}
\email{giacomo.mazza@unige.ch}
\affiliation{Department of Quantum Matter Physics, University of Geneva, Quai Ernest-Ansermet 24, 1211 Geneva, Switzerland}
\author{Adriano Amaricci}
\affiliation{CNR-IOM DEMOCRITOS, Istituto Officina dei Materiali,
Consiglio Nazionale delle Ricerche, Via Bonomea 265, I-34136 Trieste, Italy}
\author{Massimo Capone}
\affiliation{Scuola Internazionale Superiore di Studi Avanzati (SISSA),  
Via Bonomea 265, 34136 Trieste, Italy}
\affiliation{CNR-IOM DEMOCRITOS, Istituto Officina dei Materiali,
Consiglio Nazionale delle Ricerche, Via Bonomea 265, I-34136 Trieste, Italy}

\begin{abstract}
 We consider heterostructures obtained by  stacking layers of two s-wave superconductors
with significantly different coupling strengths, respectively in the weak- and strong-coupling regimes.
The weak- and strong-coupling superconductors
are chosen  with similar critical temperatures for bulk systems.
Using dynamical mean-field theory methods, we find an ubiquitous enhancement of the superconducting 
critical temperature for all the heterostructures
where a single layer of one of the two superconductors is alternated with a thicker multilayer of the other.
Two distinct physical regimes can be identified as a function of the thickness of the larger layer: 
(i)  an inherently inhomogeneous superconductor characterized by the properties of the 
  two isolated bulk superconductors where the enhancement of the critical temperature
  is confined to the interface and (ii) a bulk superconductor with an enhanced critical 
  temperature extending to the whole heterostructure.  We characterize the crossover between these regimes in terms of the competition between two length scales 
connected with the proximity effect and the pair coherence.  
\end{abstract}

\maketitle
\section{Introduction}
The design of artificial heterostructures has established as an ideal framework to
engineer the properties of functional materials by combining materials with different bulk behavior.
The advances in our ability to control the properties of heterostructures allows to realize tunable quantum phenomena, 
as it happens in the spectacular example of twisted blayer graphene~\cite{ycao_sc_tblg,yankowitz_tblg}.
Remarkable examples include heterostructures based on 
oxides~\cite{tokura_emergent_phenomena_oxide_interface,
zubko_annual_review2011,caviglia_interfaceSC_LAO_STO,bozovic_2008}
or  two-dimensional Van der Walls materials~\cite{geim_grigorieva_vdw_heterostructures}.

Any heterostructure is built by a series of interfaces between different materials where a variety 
of fascinating phenomena such as magnetisms~\cite{magnetism_interface_RMP,gibertini_2d_magnets}
or superconductivity~\cite{caviglia_interfaceSC_LAO_STO,bozovic_2008,ddicastro_CCO_STO} can be observed
even when they are not present in the bulk of the constituent materials.
The periodic  repetition of interfaces in different patterns offers a further handle to design
and engineer artificial compounds thereby 
controlling their functional properties~\cite{claribel_NNO_SNO}.

Superconducting materials are among the most used bricks to build heterostructures, not only for their
intrinsic interest, but also because they are known to show the proximity effect, which is associated with Cooper pairs leaking 
from a superconductor to another material across an
interface.~\cite{de_gennes_boundary_sc,bozovic_gpe,cherkez_ss_proximity_prx}
For example, the proximity effect can be used to stabilize a superconducting state at the
interface, which either does not exist in the bulk compound or it requires different conditions to be realized.

This has been proposed and realized in seminal works, mainly using cuprate high-temperature superconductors  .~\cite{berg_kivelson_2008,bozovic_2008,okamoto_superlattice_hubbard_cdmft}
In these examples, the key is to exploit the different properties of 
under- and over-doped compounds in order to enhance the overall superconductivity.
These ideas can have a very wide range of applications, for example building heterostructures of materials
with different pairing source or pairing symmetry and/or displaying other quantum phases. 



In this work we consider heterostructures in which the two components are two s-wave superconductors characterized by a different 
strength of the superconducting coupling, which puts one of the two systems in a weak-coupling 
regime and the other in the opposite strong coupling regime. 
We choose an s-wave pairing as described by a simple attractive Hubbard model. 
We can tune  the two systems to have a similar critical temperature in the bulk owing to 
the non-monotonic behaviour of the critical temperature as a function of the coupling 
in models with tunable attractive interaction.
We consider different patterns and we study the evolution of the superconducting state using an extension of the Dynamical 
Mean Field Theory (DMFT)~\cite{Georges1996RMP,Kotliar2006RMP}  designed for layered structures~\cite{potthoff_rDMFT,Amaricci2014PRA,Mazza2016PRL,petocchi}. 
The use of DMFT allows to treat the different regimes of superconductivity without any 
perturbative assumption or bias.

Besides its fundamental character, this analysis has also a relevance for real systems, since a crossover from weak to strong coupling 
can be used as a very simple effective picture of the evolution of superconductivity moving from overdoped to underdoped cuprates~\cite{Chen2004A,Chen20051,Garg2005PRB,Toschi2005NJOP,toschi_capone_castellani_PRB2005,Ho2009PRA}, which can be realized also in  the two-dimensional Hubbard model~\cite{PhysRevB.86.241106,Fratino2016SR} and it has been recently proposed also in iron-based superconductors~\cite{Kasahara2016NC,Rinott2017SA}.


Our main result is that in the case where a single layer of one material is hosted into a thicker slab of the other,
we find an enhancement of the critical temperature  with respect to  {\it both} the isolated samples. 
We understand this result discussing how the heterostructuring cures the weaknesses of the two bulk superconductors
leading to an optimized superconducting phase which can be pictured as an effective intermediate-coupling superconductor.
 We show how such effective critical temperature
is affected by the periodicity of the heterostructure
and we rationalize  the results in terms of two length scales, 
associated respectively to the proximity effect between weak and strong coupling superconductors and
to the coherence properties of the pairs. 


The rest of this manuscript is organized as follow: in
Sec.~\ref{sec:model} we introduce the model for the heterostructure
and discuss physical its properties in different limits. In
Sec.~\ref{sec:results} we present the results concerning the critical
temperature and its evolution with some relevant model parameters. 
The discussion of the main results and their physical interpretation
in terms of phenomenological quantities is the subject of Sec.~\ref{sec:discussion}.

\section{Model}
\label{sec:model}

We model the heterostructure in terms of a simple attractive Hubbard model where every site
experiences a local interaction between two fermions on the same site
\begin{equation}
H_{int } = - \sum_i U_i n_{i \up} n_{i \down},
\end{equation}
where $n_{i \up}$ and $n_{i \down}$ are the number operators for the fermions and $U_i > 0$. 
The two different superconductors forming our heterostructure will be characterized by different values of $U_i$.

For uniform systems with constant $U$, the model has been widely investigated using several methods ~\cite{Nozieres1985,PhysRevB.69.184501,refId0,B_nemann_2005}, including DMFT~\cite{Keller2001PRL,capone_castellani_grilli_PRL2002,randeira_PRB2005,toschi_capone_castellani_PRB2005,Tagliavini,Privitera1,Privitera2}   and its
extensions~\cite{KyungGeorges,Amaricci2014PRA,petocchi,del_re_DGA_PRB2019}.

The ground state of the model is superconducting with s-wave symmetry for every value of $U$, 
but the properties of the 
superconducting state evolve in a non-trivial way as a function of the ratio between the interaction and the bandwidth $W$ of the non-interacting model.

At weak coupling, when $U$ is much smaller than $W$, the superconducting state is well described by a BCS-like theory with an instantaneous attraction
and  the superconducting state is characterized by the formation of weakly bound Cooper pairs with a large correlation length. 
Upon increasing the temperature the pairs are progressively broken, 
until we reach the  critical temperature where the pairs are completely 
destroyed and the system becomes a normal metal.

In the opposite limit of  strong attraction for $U/W \gg 1$, 
the superconducting state is associated with the formation of tightly 
bound pairs with a short correlation length and  quickly loosing
their coherence with increasing temperature. 
This is often called a Bose-Einstein condensation (BEC) regime where "preformed"
bosonic pairs condense when the critical temperature is reached from above.
In the BCS regime $U/W\ll 1$ the critical temperature $T_c$ is exponentially small
and increases as a function of the attraction. Instead, in the strong coupling
regime $T_c$ decreases as $t^2/U$.  
This results in a non-monotonous behaviour of the critical temperature
as a function of $U/W$. 
The optimal critical temperature
is realized for an  intermediate pairing strength $U\simeq W$ for which none of the two limiting scenarios applies. 
In this regime we have a delicate balance between a large pairing amplitude $\phi$ and a high mobility 
of pairs leading to coherent quantum state. 
The non-monotonic behavior of $T_c$ is realized despite the zero-temperature superconducting order parameter
$\phi = 1/N \sum_i \langle c_{i \up} c_{i \downarrow}\rangle$ measuring the amplitude of Cooper pairs
increases monotonically with $U/W$. The large value of $\phi$  in the BEC limit
reflects the fact that the electrons are tightly bound in pairs, while the low critical temperature 
follows from the reduced mobility of the pairs. At the critical temperature the
modulus of the order parameter remains finite, but the system is a collection of local pairs with disordered phase
hence the average of $\phi$ as complex number vanishes.
The dome-shaped behaviour of the critical temperature is the result of an optimization of the superconducting properties
 and it matches a common experimental trend in materials, where the critical temperature reaches maxima as a function of
 doping as in the cuprates or of other control parameters like pressure. These results
 for the homogeneous system guide us to understand the properties of the heterostructures.

\begin{figure}
\includegraphics[width=\columnwidth]{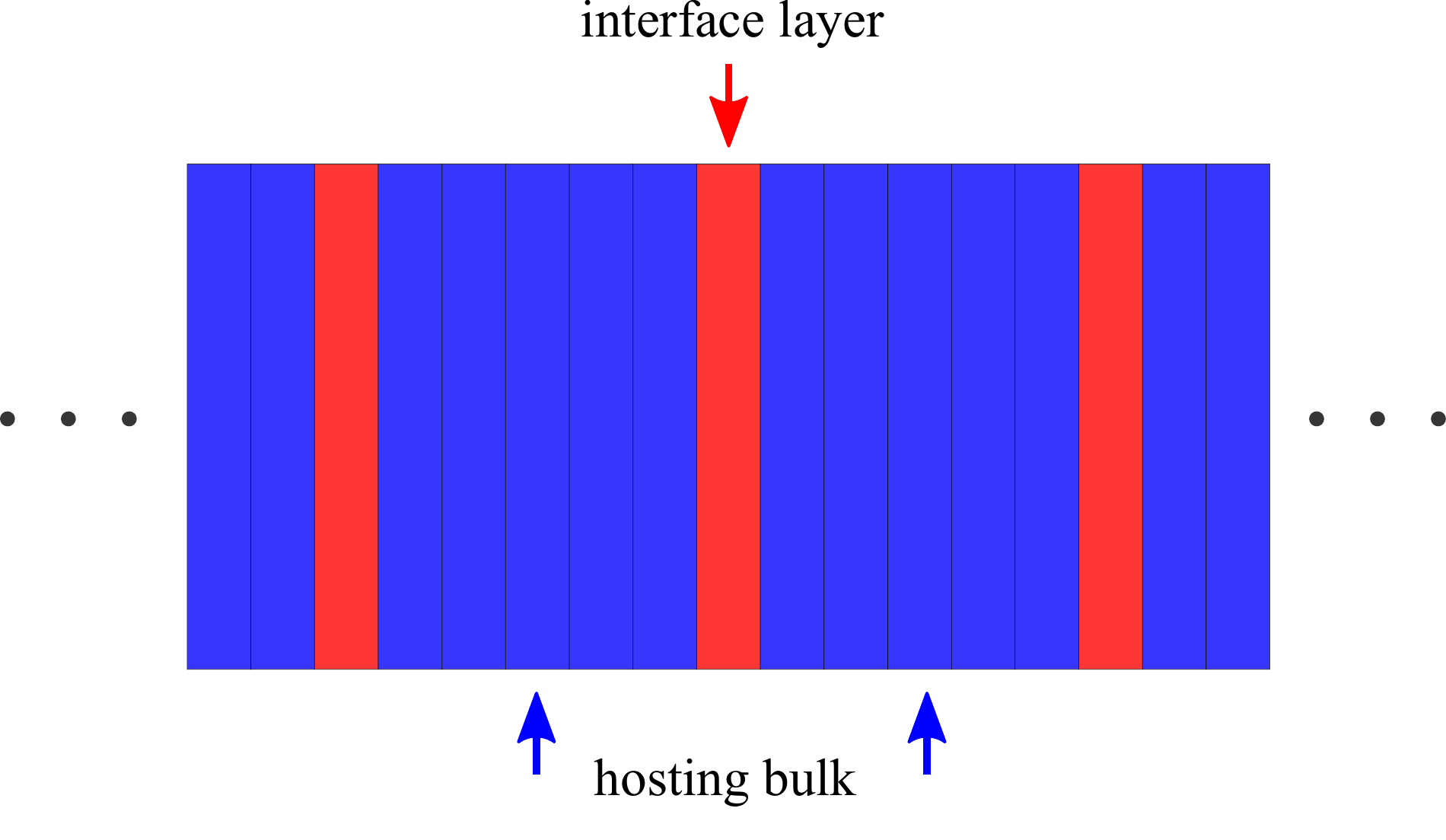}
\caption{
Sketch of  the heterostructures considered in this work.
The different colors indicate two different types of superconductors. In the following the will denote the strong-coupling 
superconductor in blue and the weak-coupling one in red.}
\label{fig_1}
\end{figure}

In the rest of this work we consider a layered structure.   
The system is assumed to be spatially homogeneous in the  
the $xy$-plane,  while it has a modulation along the $z$ direction given by the
alternate stacking of two different 
layered superconductors, respectively at weak-coupling (SC-w) and
strong-coupling (SC-s).
The periodic pattern is determined by the infinite repetition of slabs
of $N_\mathrm{w}$ (weak-coupling) and $N_\mathrm{s}$
(strong-coupling) layers. 
$(N_\mathrm{w}=1,N_\mathrm{s}=0)$ and 
$(N_\mathrm{w}=0,N_\mathrm{s}=1)$ realize two homogeneous bulk 
superconductors with weak and strong coupling characters, respectively.

Accordingly, we split the site index in terms of the 
in-plane coordinate  $\bd{R}$ and $z$ the layer index $z$. The interaction depends
only on the layer coordinate $U \equiv U(z)$. 
The layered structure Hamiltonian reads:
\begin{equation}
\begin{split}
  H(N_\mathrm{w},N_\mathrm{s}) =& 
	\sum_{\bk \sigma z} 
	\sum_{z} \epsilon(\bk) \cc_{\bk z \sigma} \ca_{\bk \parallel z \sigma}   +
	t \cc_{\bk  z \sigma} \ca_{\bk (z+1) \sigma} + h.c.\\
 	&- \sum_{ \bd{R} z} U(z) n_{\bd{R} z \up} n_{\bd{R} z \down}
 	- \mu \sum_{\bd{R} \s} n_{\bd{R} z  \s}
\end{split}
\label{eq:inhomo_hubbard}
\end{equation}
with  the interaction $U(z)$ varying periodically every $\L = N_\mathrm{w}+ N_\mathrm{s}$ layers
\begin{equation}
U(z)
=
\left\lbrace
\begin{matrix}
U_\mathrm{w} & & 1 +k\L \leq z \leq N_{\mathrm{w}} +k\L \\
U_\mathrm{s} &  &N_{\mathrm{w}} +1 +k\L  \leq z \leq \L +k\L \\
\end{matrix}
\right.
\end{equation}
where $k$ is an integer number.

Throughout the rest of the paper we set $U_{\mathrm{w}}/W = 0.28$ and $U_{\mathrm{s}}/W = 3.16$
which in the homogeneous system are characterized by comparable 
critical temperatures $T_{c,\mathrm{w}}/W \approx 0.015$ and $T_{c,\mathrm{s}}/W \approx 0.014$. 
In the rest of the paper we indicate with $T_{c,0} = \left( T_{c,\mathrm{w}}+T_{c,\mathrm{s}} \right) / 2$
the average of the critical temperatures of the two bulk superconductors.

Among all the possible $(N_\mathrm{w},N_\mathrm{s})$ configurations 
of the heterostructure, we consider the two extreme cases
$(1,N_\mathrm{s})$ and $(N_\mathrm{w},1)$, corresponding to 
the periodic insertions of a single layer of one type of 
superconductor into the bulk of the other. 
In the following we will refer to the periodic insertion as 
the {\it interface} and to the hosting bulk as the {\it bulk} (see Fig.~\ref{fig_1}).
Given this choice of the configuration, we introduce a single
parameter $\D N \equiv N_\mathrm{s} - N_{\mathrm{w}}$ univocally identify the heterostructure.
For $\D N > 0$  the strong coupling superconductor 
assumes the role of bulk and the weak coupling is the interface, whereas 
the opposite holds for $\D N < 0$.
The case $\D N=0$ corresponds to the 
$(\ldots - \mathrm{w}-\mathrm{s}-\mathrm{w}-\mathrm{s} - \ldots)$
heterostructure and there is no distinction between interface and bulk. 

The inhomogeneous superconducting phases of model (\ref{eq:inhomo_hubbard}) are described by using 
a real-space extension of DMFT~\cite{potthoff_rDMFT,yAmaricci2014PRA,Mazza2016PRL,petocchi} where  the self-energy 
is local in space, i.e.,  $\Sigma_{ij} = \Sigma_i \delta_{ij}$ where i and j are two lattice sites, but it can depend on the site.
In this work we explicitly enforce
translational invariance within each layer, while the different layers have a different local 
self-energy with normal and
an anomalous superconducting components $\Sigma(z)$ and $S(z)$
with the same periodicity of the heterostructure 
$\Sigma(z+\L) = \Sigma(z)$ and $S(z+\L) = S(z)$.
We solve the
quantum impurity models associated to each layer using a Lanczos-based exact diagonalisation algorithm
at zero and finite temperature.~\cite{Capone2007PRB,Weber2012PRB}.
Finally, we impose a half-filling condition for every layer. Therefore  we inhibit charge-density wave solutions, which are known
to exist for the attractive Hubbard model and any process where the charge distribution is not homogeneous. This choice allows us to focus on superconductivity 
and the intrinsic effect of the heterostructuring.

\section{Results}
\label{sec:results}
\subsection{Enhancement of the critical temperature}
\label{subsec:A}
\begin{figure}
\includegraphics[width=\columnwidth]{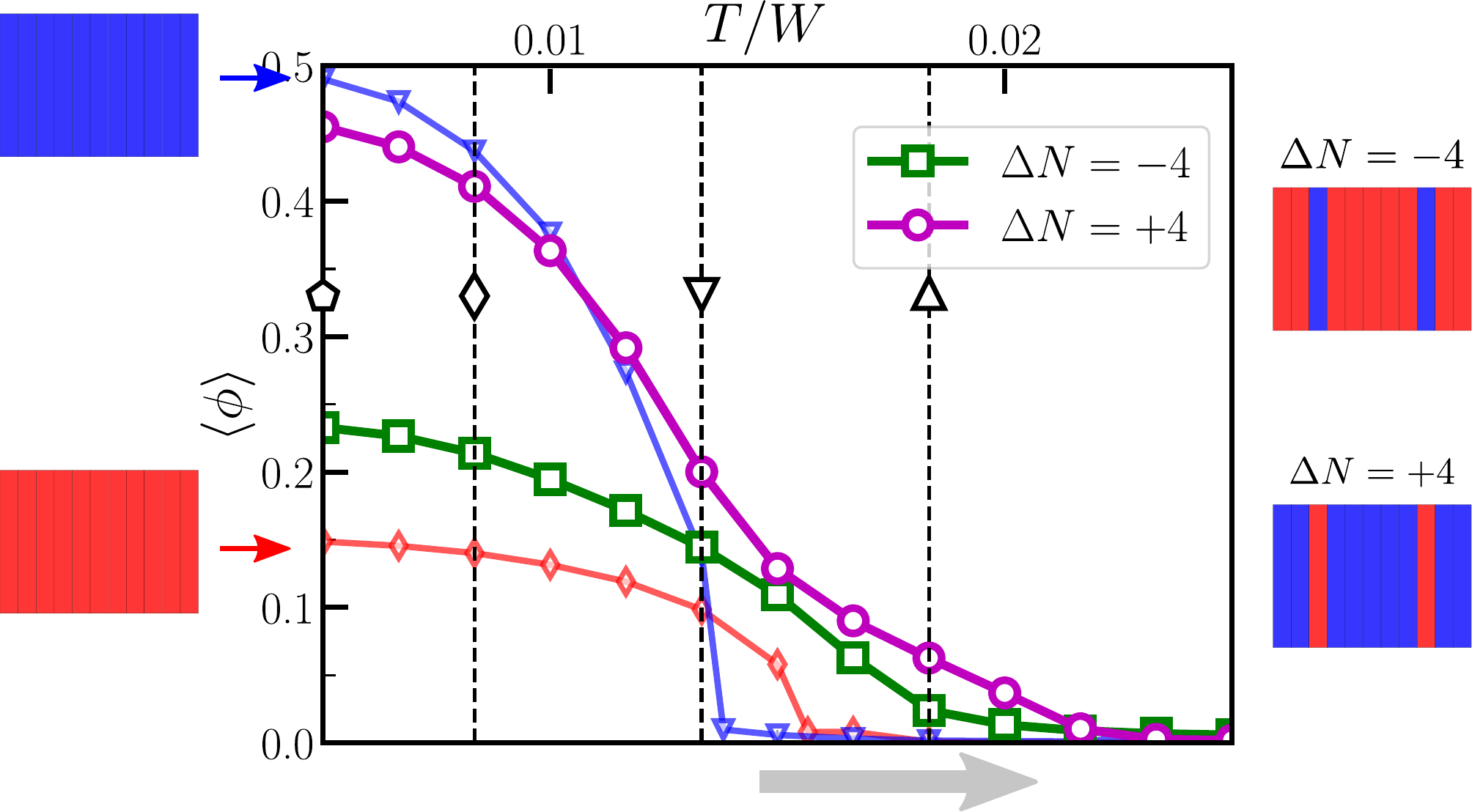}
\caption{ 
  Spatial average of the superconducting order parameter as a function of temperature.
  We compare the homogeneous cases (thin lines with small symbols) at weak
  coupling $U = U_\mathrm{w} = 0.28W$ (diamonds) and strong
  coupling $U = U_\mathrm{s} = 3.16W$ (triangles) with the heterostructures 
 with $\D N = +4$ (circles) and $\D N=-4$ (squares), whose data are connected by 
  thick lines.  
  The grey arrow  below the horizontal axis highlights the critical temperature enhancement
  with respect to the value of the homogeneous case.
  The vertical dashed lines with symbols indicate the temperature values
  used in the next Fig.\ref{fig_3}. 
}
\label{fig_2}
\end{figure}

\begin{figure*}
\includegraphics[width=0.95\textwidth]{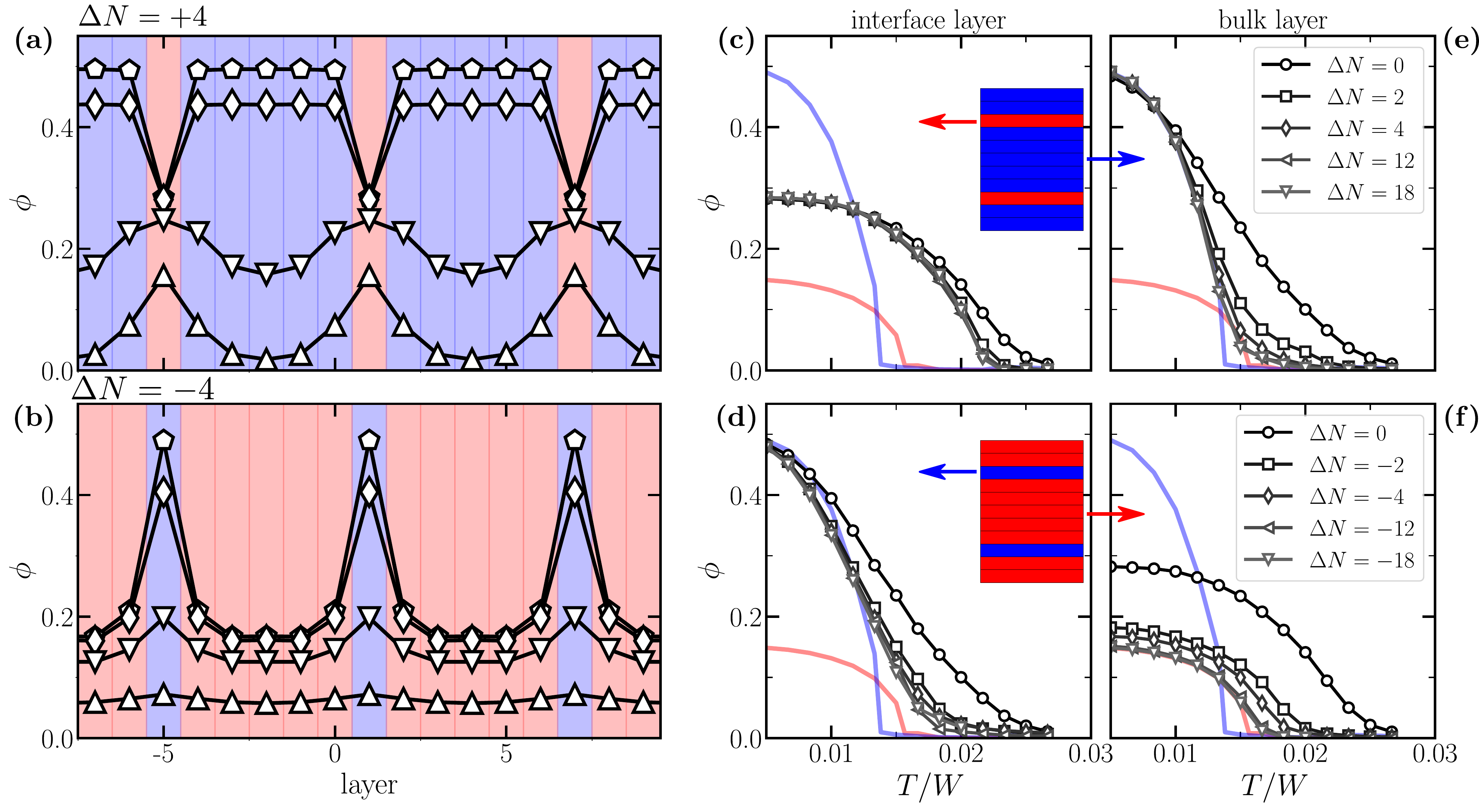}
\caption{
Left Panels (a-b):
  Order parameter as a function of the layer index at different temperatures 
  for $\D N= + 4$ (top) and $\D N=-4$ (bottom). 
  The temperatures of the order parameters profiles are indicated by 
  vertical lines with the corresponding markers in Fig.~\ref{fig_2}.
  Right Panels (c-f):
  Temperature evolution of the order parameter on the interface (c-d)  and in the bulk (e-f). 
	Panels (c) and (e) refer to the case $\D N \geq 0$ whereas 
	panels (d) and (f) refer to the case $\D N \leq 0$.
	The $\D N =0 $ case is shown for both configurations.
}
\label{fig_3}
\end{figure*}

We start the discussion highlighting the main result of this work, namely the overall enhancement of the 
critical temperature of the heterostructure with respect to the constituents. 
We illustrate this enhancement for the two cases $\D N=\pm 4$ in Fig.~\ref{fig_2}. 
For each layer we compute the local superconducting order parameter 
$\phi(z) = \langle c_{\bd{R} z \up} c^{\dagger}_{\bd{R} z \up}\rangle$ 
and we compute the average along $z$, $\langle \phi \rangle = \frac{1}{N} \sum_z \phi(z)$.

Notice that the two homogeneous superconductors have fairly different zero-temperature order parameter despite their very close critical temperatures. 
While the zero-temperature order parameters of the two heterostructures fall between the two homogeneous systems, we observe that the average order parameter 
clearly remains non-zero up to temperatures significantly larger than the homogeneous $T_c$. 
 
In the homogeneous case the order parameter falls to zero with a rather sharp behaviour compatible with
a square-root  behaviour $\phi^2 \sim 1-T^2/T_c^2$.
On the other hand, the decrease in the heterostructure as a function of temperature is much smoother, mainly as a consequence of  the inhomogeneous nature of the order parameter. 
 

A first simple interpretation of these results can be drawn in terms of a proximity 
effects which has already been discussed at interfaces between s-wave superconductors\cite{petocchi}.
The key observation is that, for bulk superconductor, superconducting order requires finite
values of the complex order parameter, which in turn requires both a finite pairing amplitude and a fixed value of the phase. 
The latter is associated to the coherence between the pairs, which is ultimately related to the pair mobility. The critical temperature is 
basically set by the "weak-link" between the two, i.e., the condition which is harder to meet.
More concretely, the weak-coupling superconductor is characterized by a coherent motion, but the critical 
temperature is low because of the small pairing amplitude, while the  strong-coupling system has a large pairing amplitude, but the pair mobility and the
relative coherence are small (proportional to $t^2/U$), leading to a critical temperature 
decreasing as the ratio $U/W$ grows. 

In the heterostructure proximity effects are expected to compensate for the weaknesses of the two bulk superconductors.
The proximity of the BEC superconductor can enhance the pairing amplitude 
on the weak coupling side which can be pictured as a leaking of Cooper pairs. 
On the other hand, the  strong coupling side can benefit from an enhancement of the mobility induced by
the coupling with BCS layers, similar to what happens 
in heterostructures involving metals and Mott insulators,
where the quasiparticles at the metallic side increase their degree of localization through coupling with the Mott insulator~\cite{freericks_fragileFL_slab,Borghi_prb10,giacomo_phd_prb}.

In a nutshell, both the superconductors are supplied by the other with a boost in the weak link quantity. 
In very loose terms, we can picture the enhancement of the critical temperature as an effective 
intermediate-coupling system which realizes artificially the ideal conditions for superconductivity. 
In the following we explore in more details the results
to make the physical picture more concrete and definite.




\subsection{Inhomogeneous superconductivity: Interface and bulk order parameters}

We now investigate  in details the inhomogeneous character of the
superconductivity in the heterostructure. In Fig.~\ref{fig_3}(a-b)
we report  the order parameter profile as a function  of the layer index for the 
two cases $\D N=\pm 4$  introduced above for the four temperatures marked in Fig. 1 
with symbols. 
At  low temperatures (pentagons and diamonds, in the order) the profiles qualitatively follow
the weak- and strong-coupling characters of each layer, showing alternated minima and maxima. 
However, upon increasing the temperature the order parameter profile evolves
following completely different behaviours in the $\D N=4$  and $\D N=-4$ configurations.
 
For $\D N=4$ (panel (a), 1 layer of weak-coupling and 5 of strong-coupling) 
the large values of the order parameter found in the strong coupling section 
are rapidly suppressed with temperature.  
Close to the critical temperature superconductivity remains thus mostly 
confined to the weak coupling interface, whose order parameter decreases much more slowly. 

On the other hand the temperature evolution in the 
$\D N=-4$ (panel (b), 1 layer of strong-coupling and 5 of weak-coupling) case is
more homogeneous in the different layers. In this case the strong and weak coupling layers
progressively reduce the order parameter, though the BEC interfaces
undergo a faster decay. The striking difference is that in this case, near the critical temperature the system is
characterized by a quasi-homogeneous superconducting order parameter.

In order to gain further insight about the mechanism leading to the enhancement of 
critical temperature and to the different scenarios we have discussed, we consider
the behaviour as a function of the number of layers in the bulk section, parameterized by $\D N$.
For simplicity we focus on  the order parameters at the  interface and at the
central layer of the hosting bulk, hereafter indicated as the bulk layer. 
In Fig.~\ref{fig_3}(c)-(f) we report the temperature 
evolution of the interfaces (panels (c) and (d)) and 
bulk layers (panels (e) and (f)) for different 
values of $\D N>0$ (panels (c) and (e)) 
and  $\D N<0$ (panels (d) and (f)).
For the $\D N=0$ case the two order parameters are found to vanish
concomitantly at a transition temperature $\sim 1.75 T_{c,0}$.
Remarkably, this corresponds to the largest enhancement  of
the critical temperature  observed in the setup considered in this
paper.
\footnote{Note that for $\D N=0$, 
where we have an alternated pattern of weak- and strong-coupling single layers, 
the distinction between interface and bulk layers becomes meaningless. 
However, to keep the notation consistent, the labelling of interface and bulk is interchanged 
going from the top to the bottom panels.}

As $|\D N|$ increases bulk and interface become more and more different.
For $\D N > 0$ (weak-coupling interface in a strong-coupling bulk, top panels) the order parameter of the interface layer weakly depends on $\D N$ and rapidly converges
to an asymptotic value. 
The bulk layer changes more substantially and it collapses onto the corresponding 
homogeneous bulk superconductor curve for large  $\D N$. 
Yet, we observe an enhancement close to the critical temperature that it is only gradually reduced 
for larger values of $|\D N|$.

The $\D N<0$ setup (strong-coupling interface in a weak-coupling bulk, bottom panels)
varies more slowly as a function of the modulus of $\D N$, while 
the critical temperatures at which the 
two order parameters vanish are closer to each other and to the homogeneous results, in agreement with
the analysis of the previous section.
Interestingly,  owing to the extremely localized nature of the pairs in
the strong-coupling regime, in this case also the interface layer resembles the behaviour 
of the homogeneous solution for low temperature.

\subsection{Critical temperatures}

In this section we explore the relation between the behaviour of the interface layer and 
that of the bulk by comparing the interface critical temperature
$T_c^{\mathrm{interface}}$  with that of the whole heterostructure,
i.e. $T_c^{\mathrm{hetero}}$. 
The critical temperature $T_c^{\mathrm{interface}}$ is naturally defined 
as the temperature at which the interface order parameter 
vanishes~\cite{tc_error_bars}.
A definition of $T_c^{\mathrm{hetero}}$ requires to take into account 
the inhomogeneous nature of the order parameter over the 
heterostructure.  In order to avoid the influence of large local
values, we define $T_c^{\mathrm{hetero}}$ as the geometric average of the order parameter
over the whole sample, which obviously coincides with the average on the building block of 
$\L = N_\mathrm{w}+ N_\mathrm{s}$ layers which is periodically repeated
\begin{equation}
\overline{\phi} = \left[ \prod_{z=1}^{\L} \phi(z) \right]^{\frac{1}{\L}}
\end{equation}

\begin{figure}
\includegraphics[width=\columnwidth]{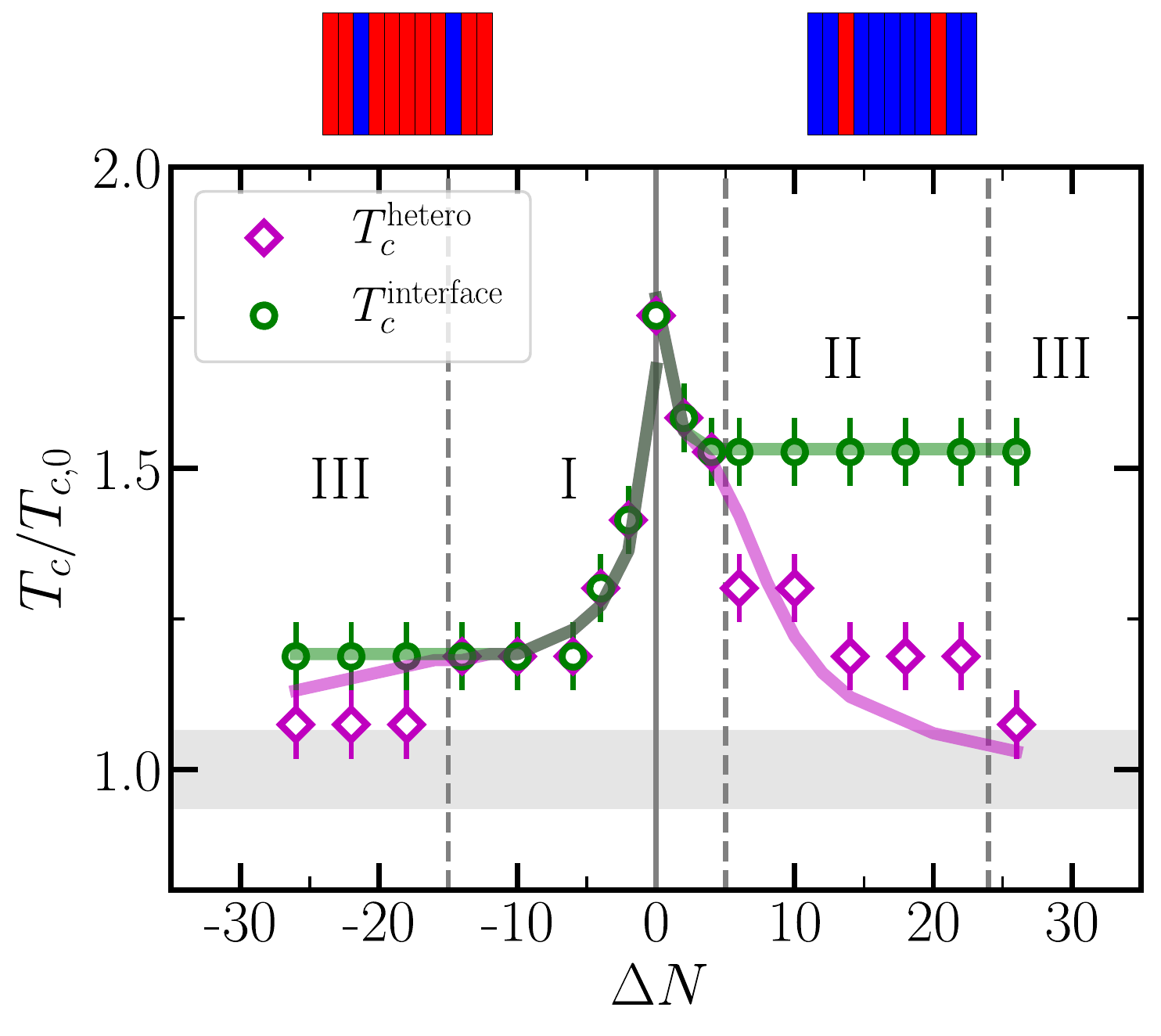}
\caption{
  Critical temperature of superconductivity confined to the interface layer $T_c^\mathrm{interface}$ 
(green circles) 
and heterostructure critical temperature $T_c^\mathrm{hetero}$ (purple diamonds) 
as a function of the parameter $\D N$.
Error bars measure the uncertainty in the determination of the critical temperature~\cite{tc_error_bars}.
Solid lines represent the critical temperatures extracted from the fit using the model
in Eq.~\ref{eq:Landau}.
The shaded area represents the area comprised between the critical temperatures 
of the two homogeneous superconductor. The enhancement is measured with respect to 
the average of the two homogeneous critical temperatures $T_{c,0}$. 
Vertical dashed lines indicate the separation of the 
different regions I,II and III of the phase diagram.}
\label{fig_4}
\end{figure}

In Fig.~\ref{fig_4} we report the evolution of $T^{\mathrm{interface}}_c$ and $T_c^{\mathrm{hetero}}$
as a function of $\D N$. 
$T^{\mathrm{interface}}_c$  achieves its maximum value at $\D N=0$ and 
decays as a function 
of $|\D N|$ reaching a rather rapid convergence for both configurations $\D N
\gtrless 0$.
In particular for $\D N>0$ the critical temperature saturates 
already for  $\D N \gtrsim 4$. The other regime $\D N<0$  is instead characterized 
by a slower decay and almost ten layers are needed to reach a saturation. 
We stress again that the interface critical temperatures are always higher than those of the two bulk 
superconductors, and that the asymptotic values 
reached in the $\D N>0$ configuration are larger.



$T^{\mathrm{hetero}}_c$ has a similar qualitative behaviour with a maximum for  $\D N =0$ and 
a decay in both directions. Yet, while the two critical temperatures are very close for
small values of $\D N$,  increasing the thickness of the bulk $T_c^{\mathrm{hetero}}$ decays below the
saturated values of the interface. This effect is particularly 
pronounced on the $\D N > 0$ side, where, as
we discussed above, the bulk layers gradually recover the properties associated with their local coupling 
strength.


Comparing the interface and the heterostructure critical temperatures 
we can identify three different regimes (marked in the figure as I, II and III)
as a function of the spacing $\D N$ between two successive interfaces.

The regime I, centered around $\D N = 0$
is characterized by a rather uniform superconducting state with enhanced critical temperature 
$T_c^{\mathrm{interface}} \approx T^{\mathrm{hetero}}_c > T_c^{\mathrm{homo}}$. 
In the other two regimes  the critical temperature of the interface is higher than that of the whole 
heterostructure $T_c^{\mathrm{interface}} > T^{\mathrm{hetero}}_c$, but in the intermediate regime II  they are both 
larger than $T_c^{\mathrm{homo}}$, while in
regime III the heterostructure converges to the homogeneous 
superconductor $T^{\mathrm{hetero}}_c \approx T_c^{\mathrm{homo}}$.

The three regimes highlight a crossover from a superconductivity enhancement which 
for large heterostructure periodicity (III) remains confined to the interfaces, while the bulk reproduces the 
result of the corresponding homogeneous material, to a system in which the  enhancement of $T_c$ 
extends to the whole heterostructure (I) when $\vert\D N\vert$ is reduced.
The evolution happens through an intermediate regime II, in which an enhancement of $T_c$ for the whole structure
is observed but the superconducting state is highly inhomogeneous with important differences between the interface and bulk critical temperatures.
It is apparent that the regime II is essentially absent for 
$\D N < 0$ in agreement with the results we have shown in the previous sections. 

\subsection{Two-length scale model}
In this section we complement the DMFT analysis with a phenomenological Landau model
which includes the specific features of our geometry and is based on two length scales that 
characterize the heterostructure superconducting state.
We start by modelling a single interface layer embedded 
into the bulk of the other superconductor.
We consider a Landau expansion of the free energy, assuming a continuous order parameter which 
varies along the z direction $\phi(z)$ and a dependence of z of the coefficient of the quadratic term.
For an interface placed at $z=z_0$ we can write
\begin{equation}
\begin{split}
F_{z_0} [\phi] & = \int dz f(z-z_0)  \\ & = \int dz \a(z-z_0) \phi^2(z) + \b \phi^4(z) + \xi_z^2 (\nabla \phi)^2.
\end{split}
\label{eq:landau_single}
\end{equation}
All the other parameters do not depend on $z$.

The key feature of the model is the definition of the 
free-energy in terms of two length scales $\ell_\mathrm{p}$ and $\xi_z$ :	
\begin{itemize}
\item{
$\ell_\mathrm{p}$
describes the spatial extent 
over which the proximity effect responsible of the critical temperature enhancement 
is active.
This effect is included in the free-energy by assuming 
a spatial dependence  of the quadratic term
$\a(z-z_0)$ through the relation 
$\a(z-z_0) \sim T-\widetilde{T}_c(z-z_0)$, where
$\widetilde{T}_c(z-z_0)$ is a fictitious space-dependent critical temperature
which is maximum at the interface and decay to the homogeneous value in the bulk
\begin{equation}
\widetilde{T}_c(z-z_0) = T_0 (1-e^{-|z-z_0|/\ell_\mathrm{p}}) + T_i e^{-|z-z_0|/\ell_\mathrm{p}}.
\label{eq:tcz}
\end{equation}
In the last equation $T_0$ represents the critical temperature of the bulk while 
$T_i > T_0$  is a parameter controlling the critical temperature enhancement 
at the interface.
}

\item{The coherence 
length $\xi_z$ which, as evident from Eq. (\ref{eq:landau_single}) has the standard relation with the energetic cost to have a spatial variation of the order 
parameter. The system therefore tends to remain spatially uniform over a length  $\xi_z$.
}
\end{itemize}

The critical temperature profile $T_c(z)$ is determined by the competition
between the two length scales. We notice that, in general, this will be different from the 
fictitious $\widetilde{T}_c(z)$ defined in Eq.~\ref{eq:tcz}.
In particular, we expect $T_c(z) \to \widetilde{T}_c(z)$ in the limit 
$ \xi_z/\ell_\mathrm{p} \to 0$, 
whereas a finite value $\xi_z/\ell_\mathrm{p} > 0$ will result in a 
renormalization of the value of the  interface critical temperature 
with respect to  the fictitious one
$T_{c}(z=z_0) < T_i$.
Eventually, for $\xi_z/\ell_\mathrm{p} \to \infty $ the model 
describes an homogeneous superconductor with $T_c(z) \to T_0.$

Starting from the single interface free energy, we define the free-energy of the 
heterostructure with periodicity $\L$ as 
\begin{equation}
\begin{split}
F_\mathrm{hetero} [\phi] &= \int dz \Phi(z),~\quad
 \Phi(z) = \sum_{n=-\infty}^{n=\infty} f(z-n \L),
\end{split}
\label{eq:Landau}
\end{equation}
with $\Phi(z+\L) = \Phi(z)$.

The stationarity condition for the functional, i.e. $\frac{\d
  F_{\mathrm{hetero}}}{\d \phi} = 0$, 
determines the temperature evolution of the order parameter $\phi(z)$
as a function of the four parameters
$(T_i,\beta,\ell_{\mathrm{p}},\xi_z)$. 
We therefore extract the Landau parameters by fitting the critical temperatures 
$T_c^{\mathrm{interface}}$ and $T_c^{\mathrm{hetero}}$ to the data in Fig.~\ref{fig_4}.
For each configuration type of heterostructure ($\D N >0$ and $\D N <0$) a single set of parameters
is used to simultaneously fit both $T_c^{\mathrm{interface}}$ and 
$T_c^{\mathrm{hetero}}$.

The best fit, shown as solid lines in Fig.~\ref{fig_4} is obtained for $\D N > 0$ using  $T_i/T_0 \approx 2.0 $, $\ell_{\mathrm{p}} \approx 1.10 $, $\b \approx 1.5$,
and $\xi_z \approx 0.35$, while  for $\D N <0$ we obtain
$T_i/T_0 \approx 1.65 $, $\ell_{\mathrm{p}} \approx 1.3 $, $\b \approx 1.5$,
and $\xi_z \approx 0.9$.
The crossover from the different regions of the phase diagram 
is overall well captured by the fitting procedure.

Comparing the two sets of optimized parameters we observe that $T_{i}^{\D N>0} > T_{i}^{\D N <0}$, 
as expected from the  behavior of the $T_c^{\mathrm{interface}}$.
In line to what discussed above, 
due to the finite values of $\xi_z$ the  fictitious 
temperatures $T_i$ correctly overestimate the saturated values of $T_c^\mathrm{interface}.$

The extracted values of the characteristic lengths strengthen our physical picture for the heterostructure 
superconductivity and its different regimes.  In particular, the crossover from the different 
regions for $\D N \gtrless 0$ is well described in terms of the ratio $\ell_{\mathrm{\mathrm{p}}}/\xi_z$.
In the case $\D N > 0$ (strong-coupling bulk), 
$\xi_z$ is the shortest length scale. In this case $T_c^{\mathrm{interface}}$ 
saturates on a scale $\L \gtrsim \ell_{\mathrm{p}}$ and,
concomitantly, $T_c$ in the bulk starts to decay towards the
homogeneous value (region II). This can be readily understood as the
coherence length of the  bulk is short and smaller than the
proximity scale $\ell_\mathrm{p}$, so it gets quickly uncorrelated
from the interface. 
On the contrary, $\xi_z$ becomes the largest length scale for $\D N
<0$ case (weak-coupling bulk), favouring 
the formation of a homogeneous superconducting state.
As a result, $T_c^{\mathrm{interface}} $ follows the behaviour of $T_c^{\mathrm{hetero}}$ for 
a wide range of periodicity, reaching a saturated value $T_c^{\mathrm{interface}}$, for $\L \gg
\ell_{\mathrm{proximity}}$, smaller with respect to the $\D N >0$ case.

\section{Conclusions}
\label{sec:discussion}
In this work we have studied by means of inhomogeneous DMFT
 the superconducting properties of a hybrid heterostructure
obtained by arranging superconducting layers with  weak and strong
coupling through the periodic intercalation of a single interface
layer of one type into the bulk of the other. 

We have show that  the  
superconducting critical temperature of the layered system is 
enhanced with respect to the critical temperatures of homogeneous superconductors with
the pairing strength of the two constituent materials.

The behaviour of the critical temperature 
as a function of the periodicity of the heterostructure 
reveals the existence of two different regimes, one in which the heterostructure 
superconductivity is dominated by the interface layer intercalated in a "bulk" of the other superconductors
and the other which is a much more homogeneous superconductor extending with similar local 
properties on the whole system.

We rationalize our results in terms of a phenomenological Landau 
model based on two length scales which control respectively the 
length scale over which the proximity effect is established 
and the coherence length of the new superconducting state which controls the homogeneity.

We can picture the heterostructure superconductivity as a state in which the system manages
to improve the properties of the two constituents. The weak-coupling superconductor is boosted
by a proximity effect in which the larger pairing amplitude of the strong-coupling system leaks, while 
the strong-coupling superconductor increases its critical temperature because the carriers become
more mobile and coherent. In a sense, we optimize the superconducting property in a similar way
as it happens for the intermediate coupling superconductor which maximizes the critical temperature
in the homogeneous system. By means of the heterostructuring the maximum critical temperature can
be obtained controlling the number of layers of the two materials,  without a fine tuning of the coupling strength. 

This effect is due to the formation, close to
the interface, of superconducting pairs with mixed
weak and strong coupling characters, realizing an
effective intermediate coupling regime which optimize the
condition for superconductivity.

Our results can be used to rationalize and predict the behaviour of heterostructures obtained 
combining materials characterized by different pairing properties and
comparable critical temperatures, as for example underdoped and overdoped cuprates chosen
on the two sides of the superconducting dome, if we assume that to some extent the doping
evolution of these materials can be described in terms of an effective strong-to-weak-coupling
evolution.
We can expect similar results also for 
other phases with long-range order, e.g. antiferromagnetism~\cite{PhysRevB.85.085124} 
or charge density waves, which show a similar evolution as a function of the coupling strength.

\section{Acknowledgements.}
We thank Antoine Georges, Jean-Marc Triscone, Jennifer  Fowlie and Claribel Domínguez for
useful discussions. 
G.M. acknowledges support of the FNS/SNF through an Ambizione grant. 
Part of this work has been supported from the  European Research Council (ERC-319286-QMAC).
A.A. and M.C. acknowledge support from H2020 Framework Programme, 
under ERC Advanced Grant No. 692670 FIRSTORM, from  Italian MIUR through PRIN
2015 (Prot. 2015C5SEJJ001) and PRIN2017 CEnTral (Protocol Number 20172H2SC4).

\bibliography{biblio}

\begin{thebibliography}{53}
\bibitem [{tc_()}]{tc_error_bars}%
  \BibitemOpen
  \href@noop {} {}\bibinfo {note} {We estimate the critical temperature as the
  value for which the corresponding order parameters become smaller than the
  tiny pairing amplitude we use in the first DMFT iteration. The error bar is
  simply the temperature difference $\Delta T$ between the last two acquired
  data points.}\BibitemShut {Stop}%
\end{thebibliography}

\end{document}